# Molecular dynamics simulation of nanoindentation on nanocomposite pearlite

Hadi Ghaffarian[1,2], Ali Karimi Taheri[1], Seunghwa Ryu[*2] and Keonwook Kang[*3]

[1] Department of Materials Science and Engineering, Sharif University of Technology, Tehran, Iran

[2] Department of Mechanical Engineering, Korea Advanced Institute of Science and Technology (KAIST), Daejeon, Korea 34141

[3] Department of Mechanical Engineering, Yonsei University, Seoul, Korea 03722

*Corresponding Authors: ryush@kaist.ac.kr, kwkang75@yonsei.ac.kr

**ABSTRACT-** We carry out molecular dynamics simulations of nanoindentation to investigate the effect of cementite size and temperature on the deformation behavior of nanocomposite pearlite composed of alternating ferrite and cementite layers. We find that, instead of the coherent transmission, dislocation propagates by forming a widespread plastic deformation in cementite layer. We also show that increasing temperature enhances the distribution of plastic strain in the ferrite layer, which reduces the stress acting on the cementite layer. Hence, thickening cementite layer or increasing temperature reduces the likelihood of dislocation propagation through the cementite layer. Our finding sheds a light on the mechanism of dislocation blocking by cementite layer in the pearlite.

**Introduction-** Pearlitic phase, a lamellar structure composed of alternating layers of ferrite and cementite, plays an important role in determining the mechanical properties of steels, such as toughness, strength and formability [1]. Compared to ferrite, pearlitic structure is known as a harder phase with higher strength due to the presence of cementite lamellae, and its mechanical properties are significantly affected by the cementite microstructure [2-3]. It has been found that the fine pearlite with small interlamellar spacing (~100 nm) and narrow cementite lamellae (~10 nm) shows higher ductility than coarse pearlite during plastic deformation [4-8]. Cementite layer act as a hard obstacle in front of ferrite dislocations and causes to dislocation pile up at the ferrite/cementite interface [9]. However, to the best of our knowledge, there has been no direct experimental or simulation study on the mechanism of the dislocation blockage by the cementite layer.

Nanoindentation is a mechanical test which uses an indenter with a known geometry to plunge into a specific site of the specimen by applying an increasing load [10]. It is widely used to determine the mechanical properties of thin films to clarify the effect of geometric confinement on mechanical properties [11-14]. It can also be used to elucidate the cementite size effect on the deformation behavior of pearlite phase at nano scales, while complex equipment setup as well as the difficulty in preparing a sample with a desired cementite thickness makes the test costly and time-consuming. These difficulties can be circumvented by employing atomistic simulations.

MD simulation has been widely used to virtually perform nanoindentation in order to investigate the elastic and plastic deformation mechanisms [15-20]. In this study, we perform a series of nanoindentation tests of ferrite-cementite nanocomposites using MD simulations to investigate the role of cementite in blocking dislocation propagation in pearlite structure at various temperatures.

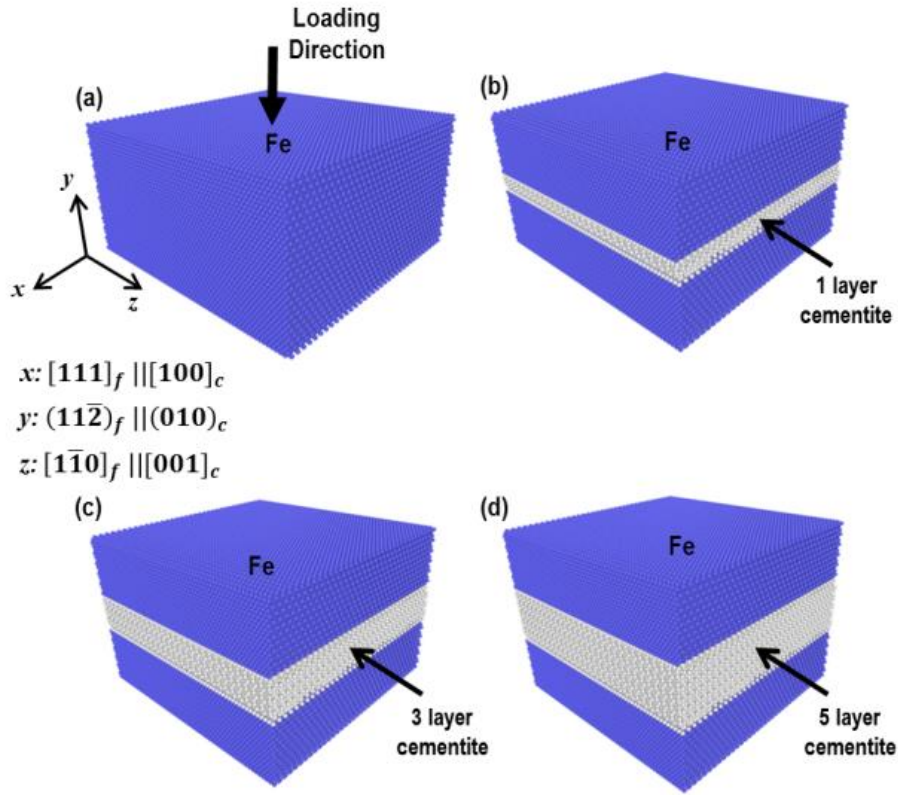

**Fig. 1.** Illustration of pearlite nanocomposite samples; (a) pure Fe, (b) P1, (c) P3 and (d) P5 samples. The position of cementite layers have been pointed by the back arrows.

**Simulation methods-** Four ferrite-cementite nanocomposite samples with different thickness of cementite ($P$=0, 1, 3, and 5 layers of cementite unit cells in the $x$-$z$ plane) were prepared as depicted in fig. 1 to investigate the cementite size effect and the temperature effect on the deformation behavior of pearlite under indentation.

We chose $[111]_f \| [100]_c$, $[\bar{1}10]_f \| [001]_c$ and $(11\bar{2})_f \| (010)_c$ for the interface orientation relationship between ferrite and cementite according to Bagaryatsky [21] (with regard to cementite

lattice constants *a*= 5.05 Å, *b*=6.69 Å and *c*=4.49 Å). Sample dimensions of pure Fe are 15 nm by 15 nm in the *x-z* plane and 8 nm in the *y* axis. The thickness of Fe layers is kept constant as 4 nm in pearlite samples, while the thickness of cementite layer is varied as 0.67, 2.0 and 3.35 nm in P1, P3 and P5 samples, respectively. The interatomic potential of Fe-C developed by Liyanage *et al*. [22], based on a modified embedded atom method (MEAM), was used to describe interatomic forces. We found that generalized stacking fault energy curves obtained from the MEAM potential are consistent with those obtained from ab initio calculations, as presented in Supplementary Materials. We also have used the potential to investigate the origin of brittle-to-ductile transition of nano-crystalline cementite [23].

The MD simulations were carried out using a parallel MD code, LAMMPS [24] with periodic boundary conditions in *x* and *z* dimensions at 100, 300 and 700 K. For each temperature, samples were relaxed for 100 ps under zero pressure with Nose-Hoover isobaric-isothermal (NPT) ensemble. All samples were indented with a constant velocity of 10 m/s along *y* direction by frictionless spherical indenter. We used three different indenter radii (*R*=3 nm, 5 nm and 7 nm) in order to investigated the effect of indenter radius on the dislocation propagation through the cementite layer. The force on each atom was calculated by:

$$f = \begin{cases} -K(r-R)^2, & r < R \\ 0, & r \geq R \end{cases} \quad (1)$$

where $K$=10 (eV/Å$^3$) is the specified force constant and *r* is the distance from the atom to the center of the indenter. To prevent translation motion during indentation, we fixed the atoms located in the region within 7 Å from the bottom of sample. The atomic arrangements of the samples, including dislocations, were then visualized using centrosymmetry parameter [25]. The atomic shear strain was also calculated using the method by Shimizu *et al*. [26].

**Result and Discussion-** Figure 2(a) shows the load-displacement curves of nanoindentation for pure Fe sample indented by 3nm radius indenter at different temperatures. We find that the indentation load gradually increases with the indenter movement up to a certain local maximum and suddenly drops due to dislocation emission from the ferrite region below the indenter tip. In fig. 2(a), the elastic responses are almost identical at different temperatures of 100 to 700 K. Our calculations show that the Young's modulus of ferrite along the loading direction does not change significantly, when temperature rises from 100 K to 700 K (see Supplementary Materials for details). This thermal insensitiveness is consistent with the experimental data that the elastic constants of single crystal Fe change only by ~10 % from 100 to 700 K [27-28]. On the other hand, the load required for the 1[st] dislocation emission decreases with the increase in temperature. Also, the increase in temperature leads to the reduction of the indentation load in plastic flow region after the first dislocation emission, and advances the emission of subsequent dislocations due to the thermal softening.

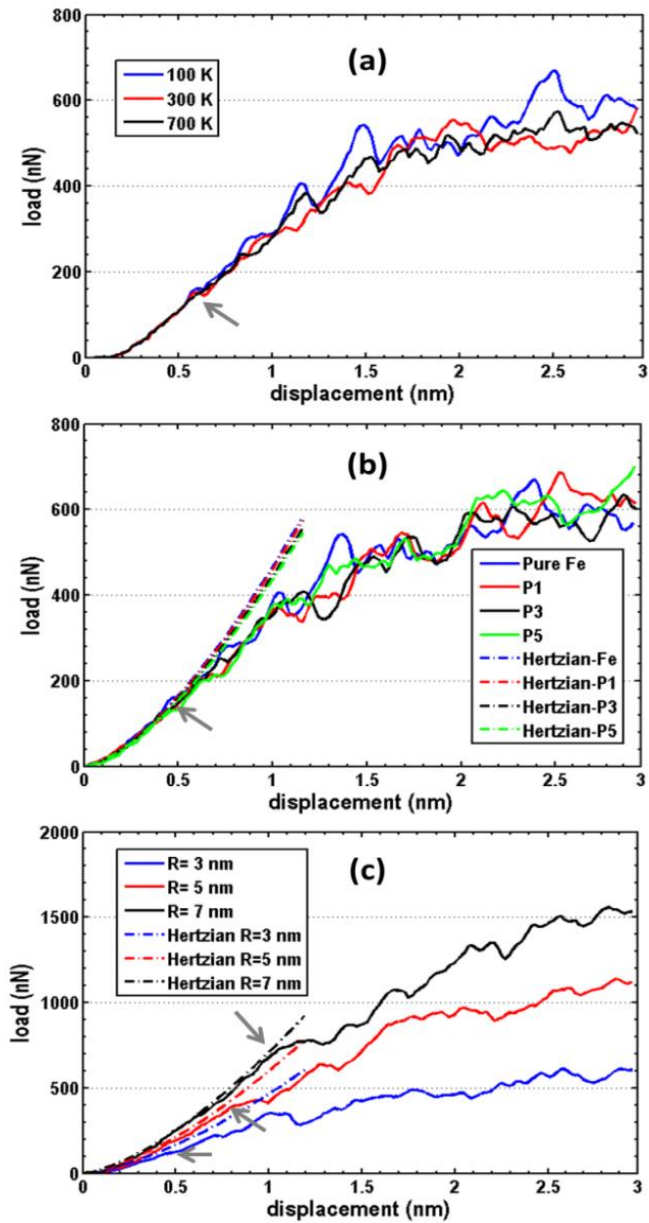

**Fig. 2.** Load-displacement curves from nanoindentation tests by indenter with 3 nm radius; (a) P0 sample at different temperatures, (b) pearlite structure with different cementite thickness at 100 K and (c) P3 sample indented by different indenter radius at 300 K. The atomistic prediction of load-displacement and Hertzian contact theory are in good agreement until the 1$^{st}$ dislocation nucleation. Arrows indicate first dislocation emissions.

The presence of cementite does not significantly affect the load-displacement curves in Fig. 2(b), because the elastic moduli of ferrite and cementite are not appreciably different [6]. We calculate the indentation load for all samples by the Hertzian theory for semi-infinite thin films,

$$F = \left(\frac{4}{3}\right) E R^{1/2} h^{3/2} \tag{2}$$

where $E$ is the Young's modulus, $R$ is indenter radius and $h$ is indentation depth. The calculated Young's moduli of ferrite and cementite along the loading direction are 203 and 159 GPa at 100 K, respectively (see Supplementary Materials for details). We use the Reuss model to estimate an effective modulus of pearlite elasticity:

$$E = \frac{1}{\frac{\rho_f}{E_f} + \frac{\rho_c}{E_c}} \tag{3}$$

where $\rho_f$ and $\rho_c$ are the volume fractions of ferrite and cementite layers in pearlite. From eq. (3), the value of effective modulus changes from 198 for P1 to 188 GPa for P5, respectively. As depicted in fig. 2(b), our MD simulation results are in good agreement with prediction from the Hertzian theory before the first dislocation emission. Similar load-displacement behaviors for P1, P3 and P5 samples are observed at 300 K and 700 K.

Figure 2(c) illustrates the load-displacement curves of the P3 sample indented by different indenter radii at 300 K. The larger indenter develops higher indentation load, due to much larger contact area. On the other hand, the increasing indenter radius causes to postpone dislocation emission due to the reduction of stress concentration below the indenter. Similar result has been reported by Ruestes *et al.* for tantalum indentation [29].

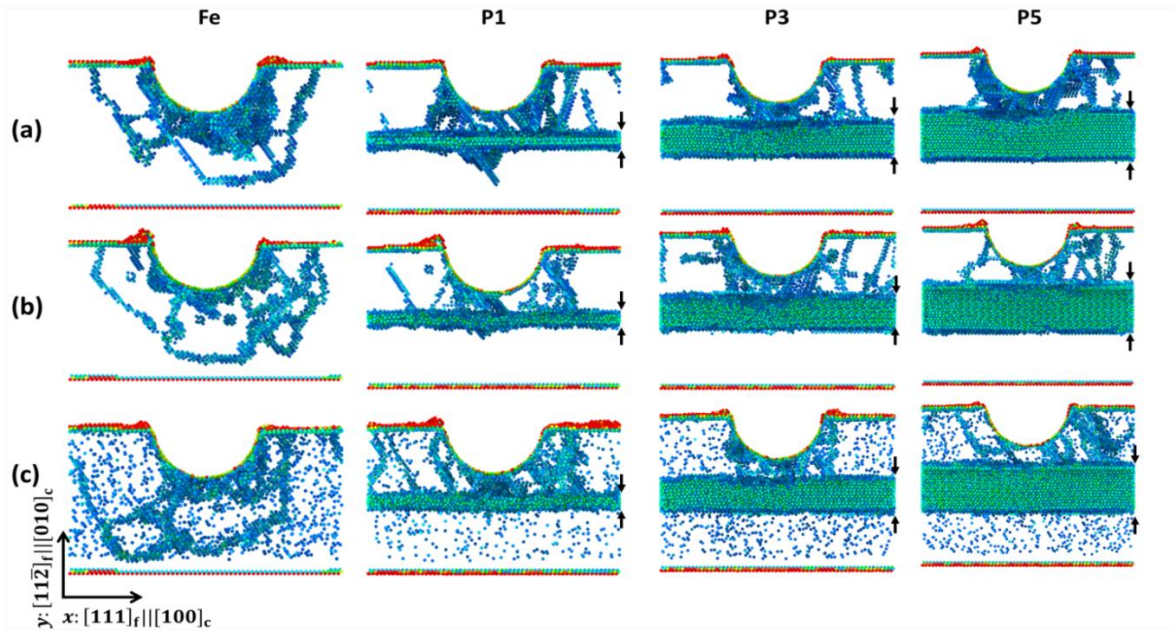

**Fig. 3.** Dislocations evolution after 30 Å indentation of indenter with 3 nm radius in pure Fe, P1, P3 and P5 samples at (a) 100 K, (b) 300 K and (c) 700 K. Out of three pearlite samples, only in P1, dislocations are formed in the bottom ferrite layer. Atoms were colored according to the centrosymmetry parameter values between 1 (blue) and 25 (red) with number of neighbors of 14. Thus, atoms on a perfect BCC lattice were not visualized. Cementite layers are illustrated by black arrows.

We present atomic configurations showing dislocation evolution after 30 Å indentation by indenter with 3 nm radius in pure Fe, P1, P3 and P5 samples at different temperatures in fig. 3, where the centrosymmetry parameter [25] was applied to screen out atoms on a perfect BCC lattice site. We observe dislocation formation in the bottom ferrite layer only in P1 sample at 100 K, while in P3 and P5 samples dislocations are completely trapped in the top ferrite layer regardless of imposed temperature.

Depending on the structural details dislocations can either transmit across the interface or be blocked at the interface. In the case of ferrite-cementite composite, dislocations could rarely pass through the incoherent bcc/orthorhombic interface of Bagaryatski's orientation relationship. The likelihood of dislocation transmission is carefully studied by investigating transmission pathway index $\chi$ [30]:

$$\chi = \begin{cases} \cos\left(\frac{\pi}{2}\frac{\theta}{\theta_c}\right)\cos\left(\frac{\pi}{2}\frac{\kappa}{\kappa_c}\right), if\ |\theta| \leq \theta_c\ ,|\kappa| \leq \kappa_c \\ 0\ , otherwise \end{cases} \quad (4)$$

where $\theta$ is the minimum angle between the intersection lines that each slip plane in ferrite and cementite makes with the interface, $\kappa$ is the minimum angle between Burgers vectors of slip systems in ferrite and cementite, and $\theta_c$ and $\kappa_c$ are limiting angles of $\theta$ and $\kappa$ for transmission. If the $\chi$ of a certain pair of slip systems is equal to or close to unity, the dislocation transmission is favorable at the interface between the two slip systems [31]. According to eq. (4), when either $\theta$ or $\kappa$ exceeds their corresponding thresholds ($\theta_c$=45° and $\kappa_c$=15° [30]), transmission is impossible. Considering main slip systems in ferrite ($\{101\}\langle11\bar{1}\rangle$ and $\{112\}\langle11\bar{1}\rangle$) and cementite ($\{100\}\langle001\rangle$), there are only 5 distinct geometrically efficient pathways with $\chi > 0.9$ among 24×6=144 possible combinations of slip system pairs (see Table 1). We find that the (110) plane is the slip plane of newly formed dislocation loops in the bottom ferrite layer of P1 sample. According to Table 1, the (110) slip plane does not belong to any possible transmission pathway on ferrite/cementite interface, and the scenario of dislocation transmission can be excluded out. Besides, the higher GSF energy in cementite [32-33] makes the dislocation nucleation in cementite harder than that in ferrite.

**Table 1.** Efficient transmission pathways for the ferrite/cementite interface of Bagaryatsky orientation relationship.

| No. | Slip system in ferrite | Slip system in cementite | $\chi$ value |
|---|---|---|---|
| 1 | $[111](1\bar{1}0)$ | | |
| 2 | $[111](\bar{1}01)$ | | |
| 3 | $[111](0\bar{1}1)$ | $[100](001)$ | 1 |
| 4 | $[111](1\bar{2}1)$ | | |
| 5 | $[111](\bar{2}11)$ | | |

The incoherent crystal structure of cementite layer acts as a barrier to dislocation transmission, which in turn develops stress concentration in cementite layer. Thus, in P1, the developed stress is high enough to deform the thin cementite layer and nucleates new dislocation loops in the bottom layer of ferrite. In contrast, in P3 and P5 samples, dislocations were not formed below the cementite layer due to smaller compliance of thicker cementite layers to transmit stress to bottom ferrite layer.

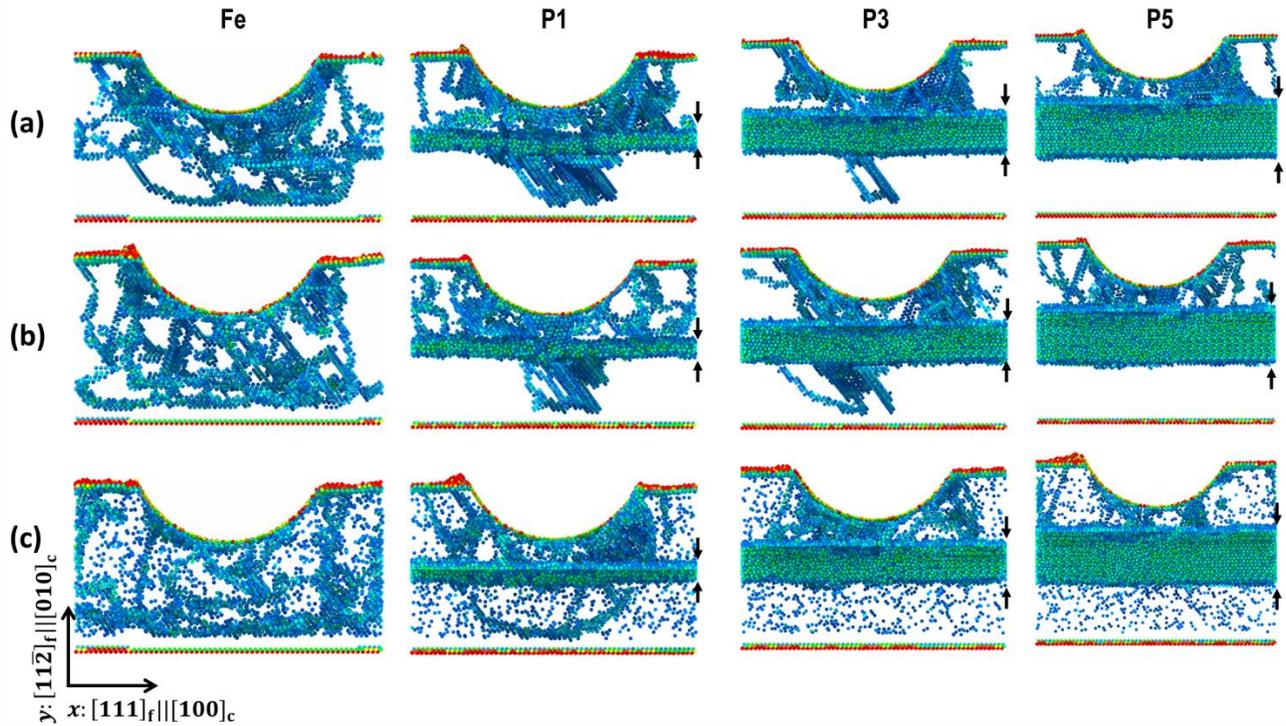

**Fig. 4**. Dislocations evolution after 30 Å indentation of indenter with 5 nm radius in pure Fe, P1, P3 and P5 samples at (a) 100 K, (b) 300 K and (c) 700 K. All conditions are same as Fig. 3.

In order to investigate the effect of indenter radius on dislocation evolution during nanoindentation, additional simulation tests were performed with 5 and 7 nm indenter radius. Figures 4 and 5 illustrate the dislocation structure in nanocomposite samples by indenters with 5 and 7 nm radius, respectively. We found higher dislocation activity in pure Fe sample when the indenter radius increases. Basically, the volume occupied by the indenter creates geometrically necessary dislocations (GNDs). The volume occupied by the indenter with 3, 5 and 7 nm radius at 30 Å indentation depth is $18\pi$, $36\pi$ and $54\pi$ nm$^3$, respectively. Hence, the larger indenter generates higher GND density at the same indentation depth. In other words, larger indenter radius induces higher stress on the cementite layer, which leads to dislocation nucleation below the thicker cementite layer. However, we find that no dislocation loops are formed in the bottom ferrite layer of P5 sample regardless of imposed temperatures during indentation by indenter with 5 nm radius, while the 7 nm radius indenter generates new dislocations below the cementite layer in all nanocomposite samples and at all temperatures. Thus, we can conclude that the dislocation blocking capability increases with the thickness of the cementite.

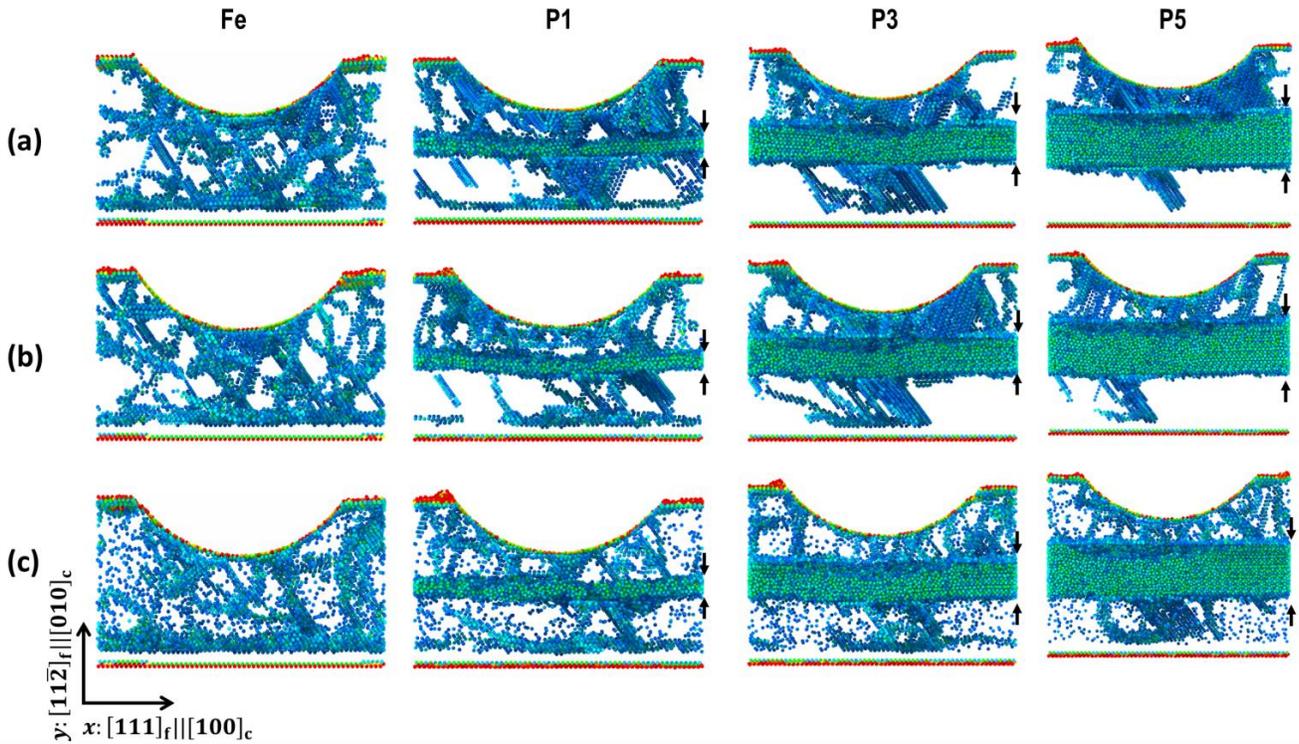

**Fig. 5**. Dislocations evolution after 30 Å indentation of indenter with 7 nm radius in pure Fe, P1, P3 and P5 samples at (a) 100 K, (b) 300 K and (c) 700 K. Dislocations are formed in the bottom ferrite layer in all deformed samples. All conditions are same as Fig. 3.

Comparison of figs. 3, 4 and 5 reveal the effect of temperature on the dislocation activity in the ferrite part. It is interesting to note that the dislocation nucleation in bottom ferrite layer decreases as temperature rises, while the activation of new slip systems are more facile in the ferrite at elevate temperature [34]. As temperature rises, the dislocations in top ferrite part (or pure Fe sample) become more distributed, and more frequently annihilate with each other. This implies that the cementite layer experiences less stress concentration, implying smaller chance for dislocation nucleation in the bottom ferrite layer.

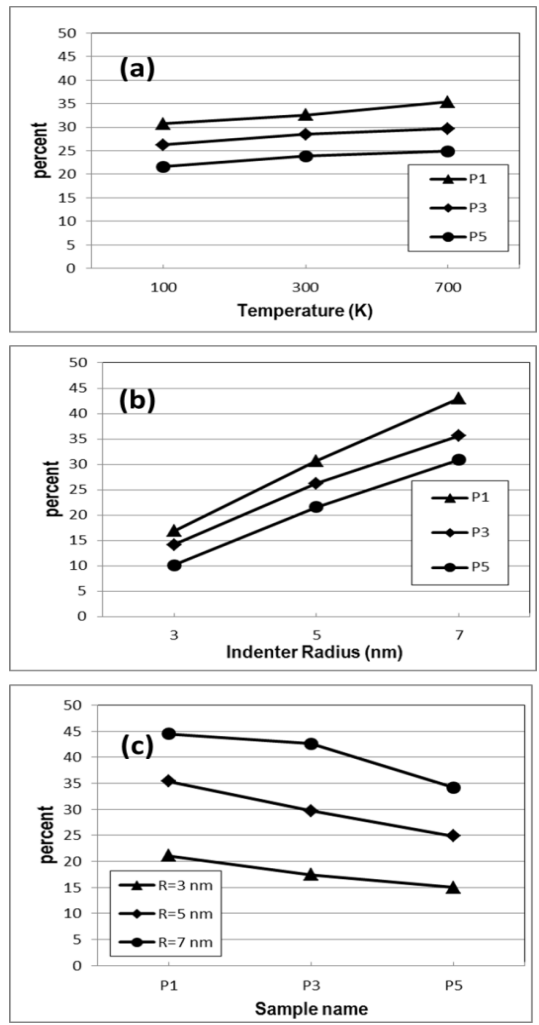

**Fig. 6**. Percent of atoms with shear strain more than 0.1 in deformed samples; (a) temperature effect, (b) indenter radius effect and (c) cementite thickness effect. The indenter radius in (a) is 5 nm and the temperature in (b) and (c) is 100 and 700 K, respectively.

In order to clarify the deformation mechanism in nanocomposite pearlite structure, the fraction of highly strained atoms in deformed samples was count following the approach by Shimizu *et al*. [26]. Figure 6(a) shows the effect of temperature on the percent of atoms with shear strain more than 0.1 in deformed samples after indentation of 30 Å by indenter with 5 nm radius (see Supplementary Materials for details of shear strain distribution during nanoindentation). As temperature rises, the percent of high strained atoms increases gradually for all deformed samples, implying more distributed deformation at elevated temperature. Same behavior was observed for other indenter radiuses.

On the other hand, fig. 6(b) reveals the effect of indenter radius on the percent of high strained atoms. An increase in indenter radius in fig. 6(b) increases the percent of high strained atoms dramatically, due to higher density of GNDs. However, we find in fig. 6(c) that the increasing cementite thickness significantly decreases the percent of high strained atoms for all indenter radii at a constant temperature. In other words, thicker cementite layer is more effective in trapping dislocation propagation and act as an obstacle for preventing plastic deformation of the bottom ferrite layer. In contrast, at lower temperature, thinner cementite layer is less likely to sustain the localized stress by the dislocation pile-up at the interface during the indentation.

**Conclusions-** In this work, we performed molecular dynamics simulations of the nanoindentation test to investigate the role of cementite in blocking dislocation propagation in pearlite structure. The load-displacement curves were found to change with temperature mainly in plastic region and the presence of cementite layer does not affect the elastic response significantly. However, the indentation load rises significantly with larger indenter radius size. We find that the cementite layer acts as a hard obstacle to dislocation propagation in pearlite. We also show that increasing temperature enhances the distribution of plastic strain in the ferrite layer, which reduces the stress acting on the cementite layer. Thus, we find that the dislocation propagation is more likely to form for a thin cementite layer at low temperature with large indenter radius.

**Acknowledgements** H.G. and A.K.T. acknowledge the Research Board of Sharif University of Technology, Tehran, Iran. K. K. and S. R. acknowledge the financial support from Basic Science Research Program through the National Research Foundation of Korea (NRF) funded by the Ministry of Education (2013R1A1A2063917) and (2013R1A1A1010091), respectively.

# Supplementary materials for

# "Molecular dynamics simulation of nanoindentation on nanocomposite pearlite"

Hadi Ghaffarian[1,2], Ali Karimi Taheri[1], Seunghwa Ryu[*2] and Keonwook Kang[*3]

[1] Department of Materials Science and Engineering, Sharif University of Technology, Tehran, Iran

[2] Department of Mechanical Engineering, Korea Advanced Institute of Science and Technology (KAIST), Daejeon, Korea 305-701

[3] Department of Mechanical Engineering, Yonsei University, Seoul, Korea 120-749

*Corresponding Authors: ryush@kaist.ac.kr, kwkang75@yonsei.ac.kr

**1. Generalized stacking fault energy of (010) plane in cementite:**

In order to test the ability of MEAM empirical potential to reproduce plasticity, dislocations, and the slip system of the cementite, we calculated generalized stacking fault (GSF) energy of the (010) plane in cementite and compared with the *ab initio* results. The (010) planes are grouped into two types; one type is a plane between two neighboring layers of only Fe atoms (labeled as type I hereinafter) while the other is the one between Fe and Fe/C layers (labeled as type II). The stacking fault (SF) energy along the [100] and [001] directions on both types of (010) planes was calculated by introducing relative slip across the slip plane in the crystal and accomplishing relaxation at T=0 K as introduced in [1]. For *ab initio* calculations, we used the Vienna ab initio simulation package (VASP) [2] with the local density approximation (LDA). A simulation cell of [100] x 3[010] x [001] containing 48 atoms (36 Fe atoms and 12 C atoms) was employed under periodic boundary conditions imposed along [100] and [001]. The *k*-points are sampled by a 4 x 1 x 4 Monkhorst-Pack method and the cut-off energy for the plane wave is 38.22 Ry. The results of molecular statics (MS) and *ab initio* simulation are shown in fig. S1. Every local minimum in fig. S1(a) indicates formation of a partial dislocation along the [100] in (010) slip planes in cementite. Both MS and *ab initio* agree that partial formation is more preferable in type I plane or a plane between only Fe-atom layers. On the other hand, in fig. S1(b), no local minima is observed along the [001] direction on type II cross section in both MS and *ab initio*. A plateau region is predicted on type I cross section by both MS and *ab initio*, and the *ab initio* curve has wider plateau with small local minima. However, at finite temperature, the effect of such a small minima would be negligible. Overall, the MEAM potential captures the characteristic features in GSF curves qualitatively well, and is considered as a reliable model to simulate plastic event in cementite. Stable stacking fault energy and the corresponding displacement are summarized in Table S1.

Besides, our results are consistent with experimental observation, where the embedded cementite particles in ferrite matrix are deformed through the movement of stacking faults, which characterized by an α[100](010) partial dislocation [3-4]. The GSF energy of (010) cementite slip plane was also calculated using ab initio method. Our results are also consistent with ab initio results, where the local minima only observed along [100] direction on both cross section of (010) cementite plane.

**Table S1.** Stable stacking fault energy $\gamma$ and corresponding displacement $\Delta x$ in the (010)[100] slip system. In MEAM MS, the lattice constants are $a$ = 5.05, $b$=6.69, and $c$=4.49 (Å) and those in ab initio are $a$ = 4.926, $b$ = 6.623, $c$ =4.377 (Å).

|            | MEAM MS       |           | Ab initio     |           |
|------------|---------------|-----------|---------------|-----------|
|            | $\gamma$ (eV/Å$^2$) | $\Delta x$ | $\gamma$ (eV/Å$^2$) | $\Delta x$ |
| Section I  | 0.079         | 0.4[100]  | 0.059         | 0.3[100]  |
| Section II | 0.096         | 0.49[100] | 0.082         | 0.3[100]  |

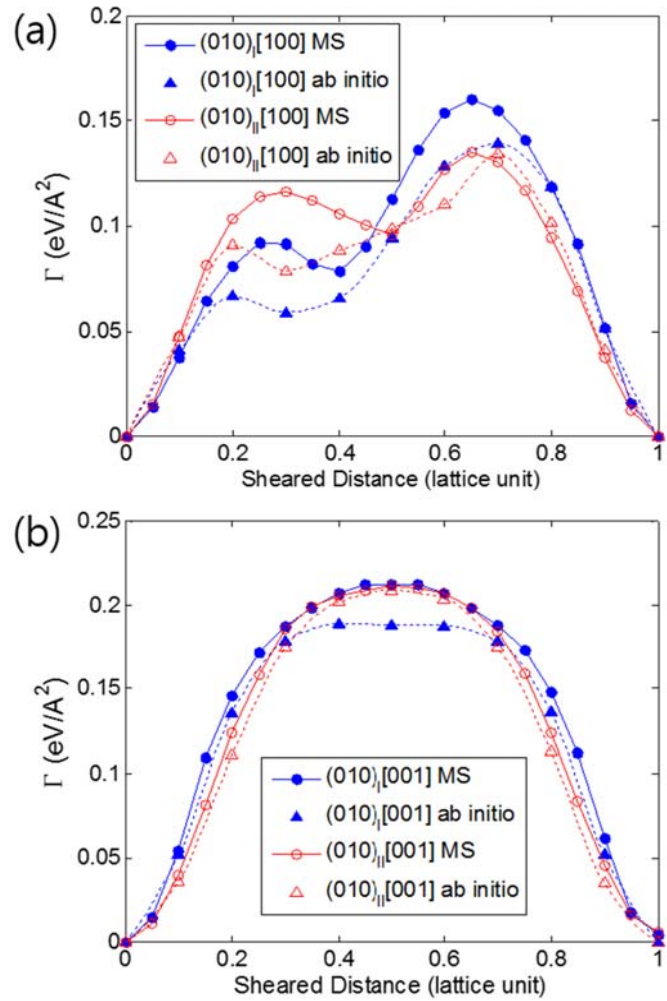

**Fig. S1.** Stacking fault energy of (010) plane along (a) [100] and (b) [001] directions with different types of cross section. Circles and triangles indicate MS results and ab initio results, respectively. The local minima indicate favorable displacement of atoms.

## 2. Calculation of elastic moduli of ferrite and cementite

In order to obtain the Young's moduli of ferrite and cementite, we calculated the elastic constants of ferrite and cementite at 100 K, 300 K and 700 K using Liyanage's MEAM potential [5]. The each component of elastic modulus tensor was computed from the stress-strain curves obtained by MD simulations at a given temperature in the NVT ensemble. Due to the complex screening function of the MEAM potential used in this study, elastic softening was observed only at the temperatures higher than 300 K. Such anomaly in elastic response would not have a considerable effect on the plastic response observed in this study.

$$C_{f-100K} = \begin{bmatrix} c_{11} & c_{12} & c_{12} & 0 & 0 & 0 \\ c_{12} & c_{11} & c_{12} & 0 & 0 & 0 \\ c_{12} & c_{12} & c_{11} & 0 & 0 & 0 \\ 0 & 0 & 0 & c_{44} & 0 & 0 \\ 0 & 0 & 0 & 0 & c_{44} & 0 \\ 0 & 0 & 0 & 0 & 0 & c_{44} \end{bmatrix} = \begin{bmatrix} 220 & 147 & 147 & 0 & 0 & 0 \\ 147 & 220 & 147 & 0 & 0 & 0 \\ 147 & 147 & 220 & 0 & 0 & 0 \\ 0 & 0 & 0 & 126 & 0 & 0 \\ 0 & 0 & 0 & 0 & 126 & 0 \\ 0 & 0 & 0 & 0 & 0 & 126 \end{bmatrix} \quad (1)$$

$$C_{c-100K} = \begin{bmatrix} c_{11} & c_{12} & c_{13} & 0 & 0 & 0 \\ c_{12} & c_{22} & c_{23} & 0 & 0 & 0 \\ c_{13} & c_{23} & c_{33} & 0 & 0 & 0 \\ 0 & 0 & 0 & c_{44} & 0 & 0 \\ 0 & 0 & 0 & 0 & c_{55} & 0 \\ 0 & 0 & 0 & 0 & 0 & c_{66} \end{bmatrix} = \begin{bmatrix} 308 & 142 & 164 & 0 & 0 & 0 \\ 142 & 233 & 119 & 0 & 0 & 0 \\ 164 & 119 & 325 & 0 & 0 & 0 \\ 0 & 0 & 0 & 24 & 0 & 0 \\ 0 & 0 & 0 & 0 & 99 & 0 \\ 0 & 0 & 0 & 0 & 0 & 57 \end{bmatrix} \quad (2)$$

$$C_{f-300K} = \begin{bmatrix} c_{11} & c_{12} & c_{12} & 0 & 0 & 0 \\ c_{12} & c_{11} & c_{12} & 0 & 0 & 0 \\ c_{12} & c_{12} & c_{11} & 0 & 0 & 0 \\ 0 & 0 & 0 & c_{44} & 0 & 0 \\ 0 & 0 & 0 & 0 & c_{44} & 0 \\ 0 & 0 & 0 & 0 & 0 & c_{44} \end{bmatrix} = \begin{bmatrix} 222 & 145 & 145 & 0 & 0 & 0 \\ 145 & 222 & 145 & 0 & 0 & 0 \\ 145 & 145 & 222 & 0 & 0 & 0 \\ 0 & 0 & 0 & 127 & 0 & 0 \\ 0 & 0 & 0 & 0 & 127 & 0 \\ 0 & 0 & 0 & 0 & 0 & 127 \end{bmatrix} \quad (3)$$

$$C_{c-300K} = \begin{bmatrix} c_{11} & c_{12} & c_{13} & 0 & 0 & 0 \\ c_{12} & c_{22} & c_{23} & 0 & 0 & 0 \\ c_{13} & c_{23} & c_{33} & 0 & 0 & 0 \\ 0 & 0 & 0 & c_{44} & 0 & 0 \\ 0 & 0 & 0 & 0 & c_{55} & 0 \\ 0 & 0 & 0 & 0 & 0 & c_{66} \end{bmatrix} = \begin{bmatrix} 280 & 145 & 158 & 0 & 0 & 0 \\ 145 & 240 & 121 & 0 & 0 & 0 \\ 158 & 121 & 308 & 0 & 0 & 0 \\ 0 & 0 & 0 & 28 & 0 & 0 \\ 0 & 0 & 0 & 0 & 96 & 0 \\ 0 & 0 & 0 & 0 & 0 & 61 \end{bmatrix} \quad (4)$$

$$C_{f-700K} = \begin{bmatrix} c_{11} & c_{12} & c_{12} & 0 & 0 & 0 \\ c_{12} & c_{11} & c_{12} & 0 & 0 & 0 \\ c_{12} & c_{12} & c_{11} & 0 & 0 & 0 \\ 0 & 0 & 0 & c_{44} & 0 & 0 \\ 0 & 0 & 0 & 0 & c_{44} & 0 \\ 0 & 0 & 0 & 0 & 0 & c_{44} \end{bmatrix} = \begin{bmatrix} 214 & 136 & 136 & 0 & 0 & 0 \\ 136 & 214 & 136 & 0 & 0 & 0 \\ 136 & 136 & 214 & 0 & 0 & 0 \\ 0 & 0 & 0 & 126 & 0 & 0 \\ 0 & 0 & 0 & 0 & 126 & 0 \\ 0 & 0 & 0 & 0 & 0 & 126 \end{bmatrix} \quad (5)$$

$$C_{c-700K} = \begin{bmatrix} c_{11} & c_{12} & c_{13} & 0 & 0 & 0 \\ c_{12} & c_{22} & c_{23} & 0 & 0 & 0 \\ c_{13} & c_{23} & c_{33} & 0 & 0 & 0 \\ 0 & 0 & 0 & c_{44} & 0 & 0 \\ 0 & 0 & 0 & 0 & c_{55} & 0 \\ 0 & 0 & 0 & 0 & 0 & c_{66} \end{bmatrix} = \begin{bmatrix} 260 & 150 & 143 & 0 & 0 & 0 \\ 150 & 244 & 112 & 0 & 0 & 0 \\ 143 & 112 & 287 & 0 & 0 & 0 \\ 0 & 0 & 0 & 29 & 0 & 0 \\ 0 & 0 & 0 & 0 & 93 & 0 \\ 0 & 0 & 0 & 0 & 0 & 59 \end{bmatrix} \quad (6)$$

Ferrite and cementite have body centered cubic (BCC) and orthorhombic structures, respectively. For arbitrary loading direction, the Young's moduli in the cubic and orthorhombic symmetry crystals are given by the following equations, respectively [6]:

$$1/E_{ferrite} = S_{11} + [S_{44} - 2(S_{11} - S_{12})](a_{11}^2 a_{12}^2 + a_{12}^2 a_{13}^2 + a_{13}^2 a_{11}^2) \quad (7)$$

$$1/E_{cementite} = S_{11} a_{11}^4 + S_{22} a_{12}^4 + S_{33} a_{13}^4 + a_{11}^2 a_{12}^2 (2S_{12} + S_{66}) \\ + a_{11}^2 a_{13}^2 (2S_{13} + S_{55}) + a_{12}^2 a_{13}^2 (2S_{23} + S_{44}) \quad (8)$$

where, $E$ is Young's modulus, $S_{ij}$ are compliance elastic constants ($\mathbf{S}=\mathbf{C}^{-1}$), and $a_{li}$ are direction cosines of the arbitrary tensile direction represented in the symmetry axes. Table S2 shows the calculated Young's moduli of ferrite and cementite along $[11\bar{2}]_f$ and $[010]_c$ at 100, 300 and 700 K:

**Table S2.** Calculated Young's moduli of ferrite and cementite along $[11\bar{2}]_f$ and $[010]_c$ at 100, 300 and 700 K.

| Temperature (K) | $E_{ferrite}$ along $[11\bar{2}]_f$ (GPa) | $E_{cementite}$ along $[010]$ (GPa) |
|---|---|---|
| 100 | 203 | 159 |
| 300 | 209 | 158 |
| 700 | 208 | 154 |

The effective elastic moduli can also be obtained from Ruess model for P0, P1, P3 and P5 samples:

**Table S3.** The effective Young's moduli of P0, P1, P3 and P5 samples along y direction at 100, 300 and 700 K.

| Sample Name | Elastic modulus at 100 K (GPa) | Elastic modulus at 300 K (GPa) | Elastic modulus at 700 K (GPa) |
|---|---|---|---|
| P0 | 203 | 209 | 208 |
| P1 | 198 | 203 | 202 |
| P3 | 192 | 196 | 194 |
| P5 | 188 | 191 | 188 |

# 3. Atomic Shear Strain Analysis

To further investigate the deformation mechanisms in detail, an atomic shear strain analysis was performed following the approach by Shimizu *et al*. [7]. Figure S2 presents the atoms with shear strain more than 0.1 in deformed samples at 30 Å indentation depth by 3 nm radius indenter at different temperatures. In pure Fe sample, shear strain spread laterally at 100 K, and increasing temperature causes more depth-wise spreading of the strained region. However, due to the existence of cementite layer, different deformation behavior is observed in P1, P3 and P5 samples, where lateral spreading of the strained region is more pronounced at elevated temperature as shown in fig. s2(b)-(c).

Figures S3 and S4 also represent the shear strain distribution in deformed sample by indenter with 5 and 7 nm radius, respectively. Comparison of shear strain distribution in figs. S2, S3 and S4 reveals the effect of indenter radius on the strain distribution in cementite layer. While the shear strain is almost trapped by the cementite layer in fig. S2 (3 nm indenter radius), the strained region spreads below the thicker cementite layer in more samples when indenter radius increases in figs. S3 and S4 due to higher dislocation activities.

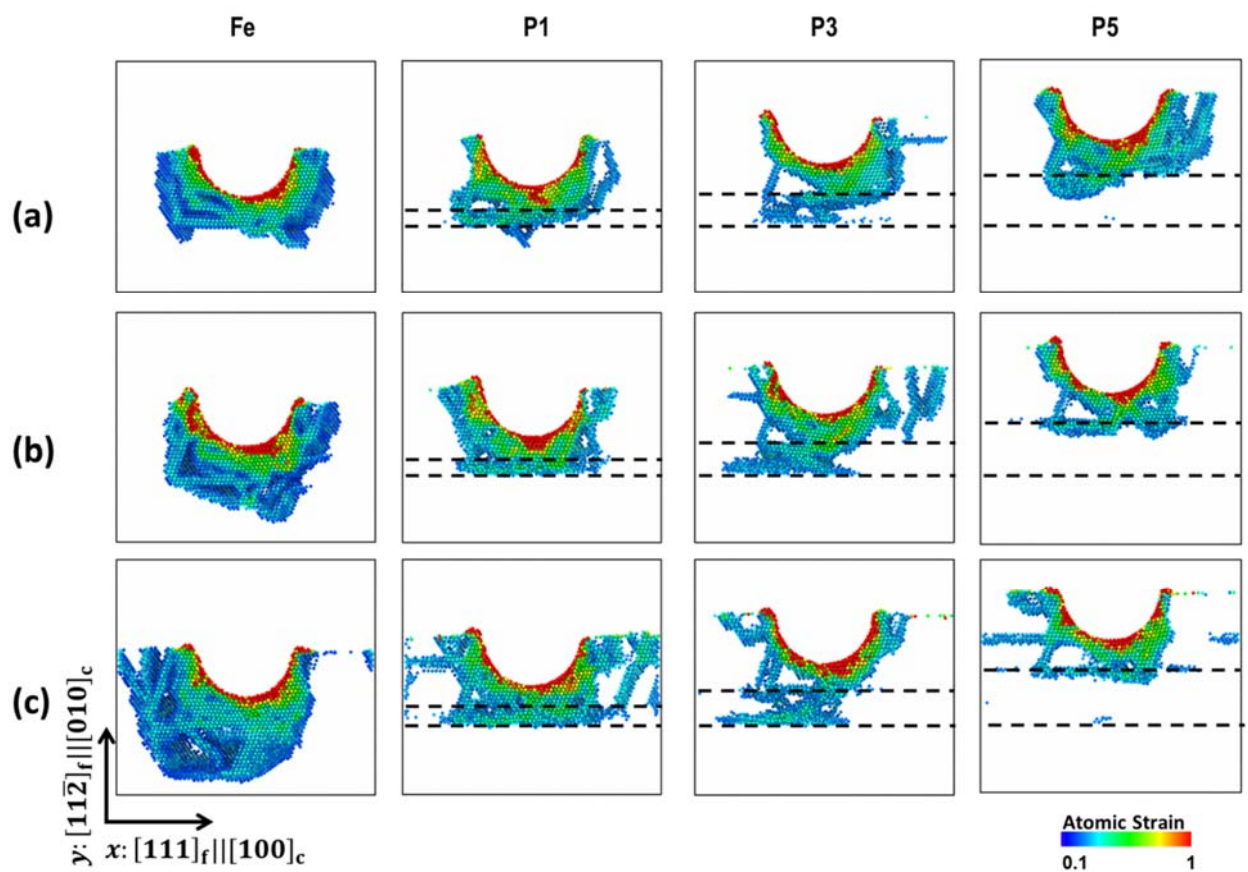

**Figure S2**. Atomic shear strain distribution of Fe and C atoms after 30 Å indentation of 3 nm indenter radius at (a) 100 K, (b) 300 K and (c) 700 K. Only atoms with shear strain more than 0.1 have been showed. Dashed lines illustrate cementite layers and red color points indicate the atoms with a shear strain equal or more than 1.

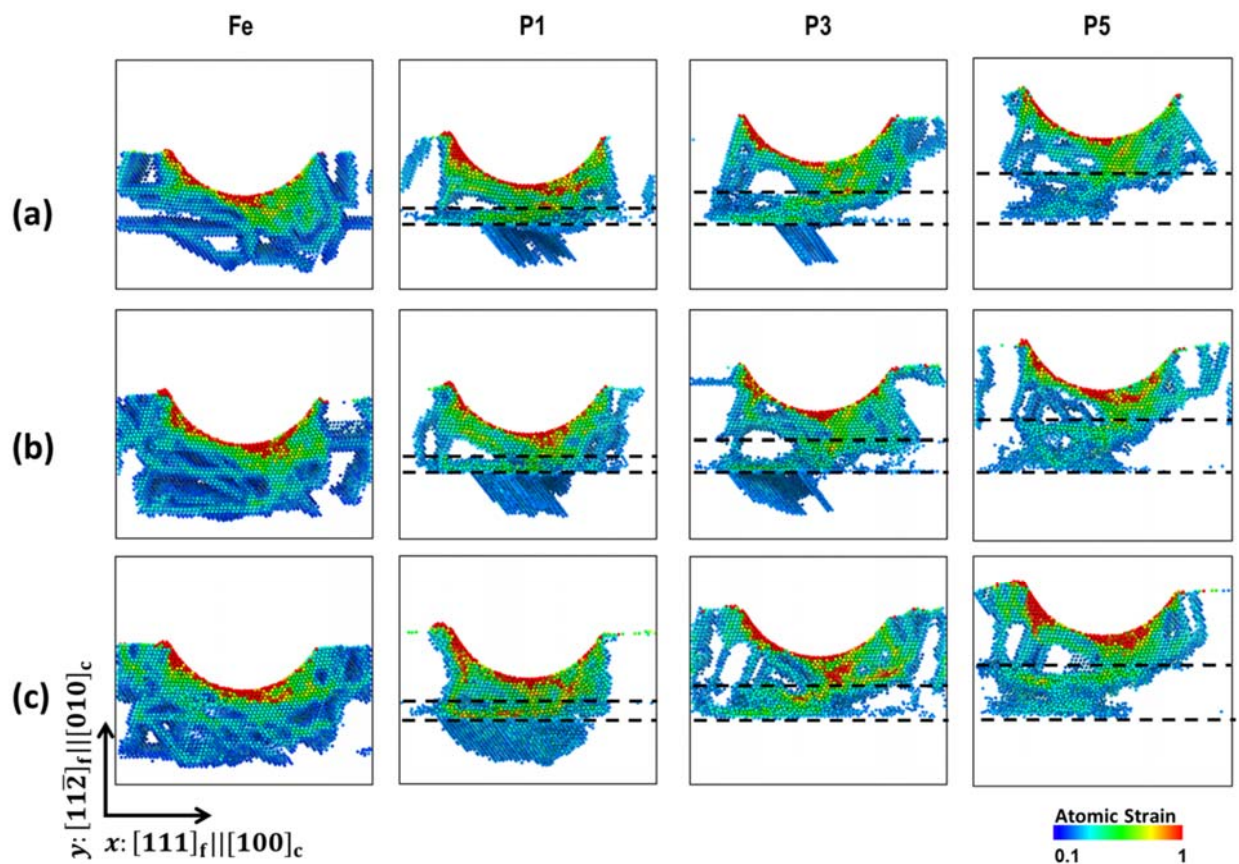

**Fig. S3**. Atomic shear strain distribution of Fe and C atoms after 30 Å indentation of 5 nm indenter radius at (a) 100 K, (b) 300 K and (c) 700 K. Only atoms with shear strain more than 0.1 have been showed.

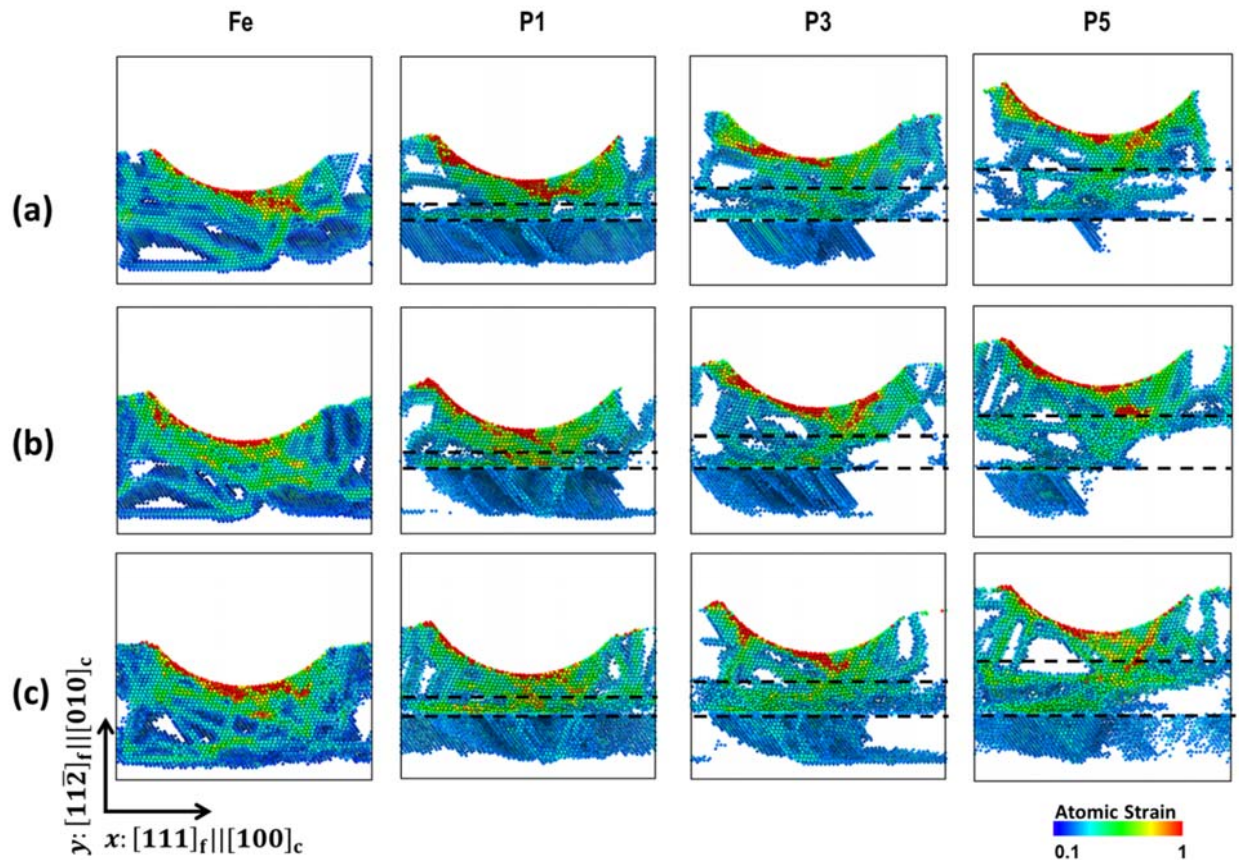

**Fig. S4**. Atomic shear strain distribution of Fe and C atoms after 30 Å indentation of 7 nm indenter radius at (a) 100 K, (b) 300 K and (c) 700 K. Only atoms with shear strain more than 0.1 have been showed.